\documentclass{PoS}
\usepackage{amssymb}
\usepackage{hep}
\usepackage{amsmath,latexsym}

\newcommand{\processgg}{\HepProcess{\Pproton \Pproton(gg) \HepTo \Ptop \APcharm + \APqt \Pcharm}}

\newcommand{\bsg}{\ensuremath{\HepProcess{\Pbottom \HepTo \Pstrange \Pphoton}}}

\newcommand{\gluglu}{\HepProcess{\Pgluon\Pgluon \HepTo \Ptop \APcharm}}
\newcommand{\mt}{m_t}
\newcommand{\tb}{\ensuremath{\tan\beta}}

\newcommand{\tcbar}{\ensuremath{\Ptop\APcharm}}
\newcommand{\ctbar}{\ensuremath{\Pcharm\APtop}}

\newcommand{\msusy}{M_{\rm SUSY}}

\title{Production of single top-quark final states at the LHC from supersymmetric FCNC interactions}

\ShortTitle{Production of single top-quark final states at the LHC from SUSY
FCNC interactions}

\author{\speaker{David L\'opez-Val} $^{a}$, Jaume Guasch $^{b,\,c}$
and Joan Sol\`a $^{a,\,c}$ \\ \, \\
        \llap{a}\, High Energy Physics Group, Dept. Estructura i Constituents de la Mat\`eria,\\
Universitat de Barcelona, Av. Diagonal 647, E-08028
    Barcelona, Catalonia, Spain \\ \\
\llap{b}\, Gravitation and Cosmology Group,  Dept. F\'isica Fonamental,\\
Universitat de Barcelona,  Av. Diagonal 647, E-08028
    Barcelona, Catalonia, Spain \\ \\
 \llap{c} Institut de Ci\`encies del Cosmos (ICC), UB, Barcelona \\ \, \\
        E-mails: \email{dlopez@ecm.ub.es},\, \email{jaume.guasch@ub.edu},\, \email{sola@ifae.es}
}

\abstract{We discuss the production of single top-quark
final states by direct supersymmetric flavor-changing interactions
at the LHC. The total cross section $\sigma(\processgg)$ is computed
at the $1$-loop order within the unconstrained MSSM. We prove that
SUSY-QCD effects may furnish sizeable production rates amounting up
to $\sim 10^5$ $\tcbar(\ctbar)$ events per $100\,\invfb$ of
integrated luminosity, in full compliance with the stringent
low-energy constraints from $\bsg$. Furthermore, we show that the
cooperative SUSY-EW effects can be sizeable on their own, regardless
of the SUSY-QCD contribution, with maximum production rates of the
order of $10^3$ events per $100\,\invfb$. Owing to the fact that
FCNC production of electrically neutral heavy-quark pairs is
virtually absent within the SM, we conclude that the observation of
such $\processgg$ processes at the LHC could lead to evidence of new
physics -- of likely supersymmetric nature.}

\FullConference{8th International Symposium on Radiative Corrections \\
                 October 1-5, 2007\\
                 Florence, Italy}

\begin{document}

\section{Introduction}
Flavor-changing neutral-current (FCNC) processes, in particular in
the context of top-quark physics, constitute a fertile ground to
seek for signatures of physics beyond the SM. The long-known
Glashow-Iliopoulos-Maiani (GIM) mechanism is nowadays in the core of
the SM, insuring the lack of tree-level FCNC interactions and also
their large supression at the $1$-loop order. Such supression does
not fully prevent us from measuring some FCNC effects, for instance
in radiative $\PB$-meson decays, $\HepProcess{\PB \HepTo X_s\,
\Pphoton}$. However, the corresponding FCNC interactions involving
the top-quark become severly inhibited. The predicted SM branching
ratios ($\mathcal{B}$) for the top-quark decays of the guise
$\HepProcess{\Ptop\HepTo\Pcharm\Pgluon}$ (${\mathcal{B}}\sim 10^{-11}$)
and $\HepProcess{\Ptop\HepTo\Pcharm\PHiggs}$ (${\mathcal{B}} \sim
10^{-14}$) are hopeless to be ever detected \cite{smtop}. Such
expected unobservability can be seen as a valuable tool for digging
into traces of distinctive new physics phenomenology.
For example, as soon as supersymmetric (SUSY) radiative
corrections are switched on, the above picture of the FCNC top-quark
dynamics may change in a dramatic way. This is the case e.g. of the
top decay into MSSM neutral Higgs bosons,
$h=\PHiggslightzero,\PHiggsheavyzero,\PHiggspszero$, their branching
ratios yielding ${\mathcal{B}}(\HepProcess{\Ptop \HepTo \Pcharm h })\sim
10^{-4}$ \cite{tdecay}, that is, $10$ orders of magnitude above the
SM prediction. Indeed, besides the usual SM charged-current
flavor-changing interactions, the MSSM flavor dynamics is also
driven by explicit intergenerational mixing terms. They stem from
the fact that the squark mass matrix does not diagonalize, in
general, within the same rotation matrix as for the quark mass one.
In order to take such misalignment into account,
a set of flavor-mixing parameters
$\delta_{ij}^{(q)\,AB}$ is introduced, 
being defined in such a way that the soft-SUSY breaking
squark mass matrix elements of type $q=u(d)$,
$i(j)$-th flavor ($i\,\neq\,j$) and $A(B)$-chirality read
$(M_{ij}^{(q)\,AB})^2=m_i\,m_j\, \delta_{ij}^{(q)\,AB}$
\footnote{We shall settle the soft-SUSY breaking
terms to a common mass scale, $m_i=M_{SUSY}\, , \,\forall i$.}.
Henceforth we will only allow non-zero mixing terms of LL type
as this possibility is the simplest one and is well motivated by
renormalization group arguments\,\cite{duncan}. The presence of
such explicit flavor-mixing parameters may provide
non-supressed sources of non-standard FCNC phenomena.

\section{Numerical results}
In what follows we shall discuss the full MSSM
contribution (including the SUSY-QCD and the
SUSY-EW effects) to $\sigma(\processgg)$, where $(\Pgluon\Pgluon)$
denote gluon-fusion. The latter is (by far) the
leading
production mechanism at the partonic level. In an obvious notation,
we have $\sigma = 2\,\sigma_{\tcbar}$. Throughout the
present work we have made use of the standard algebraic and
numerical packages \emph{FeynArts}, \emph{FormCalc} and
\emph{LoopTools} \cite{feynarts}, together with the program
\emph{HadCalc} \cite{rauch}, for the computation of the hadronic
cross sections.

\begin{figure}
\centerline{
\begin{tabular}{ccc}
\includegraphics[scale=0.45]{vai.eps} &
\hspace{-0.5cm}\includegraphics[scale=0.8]{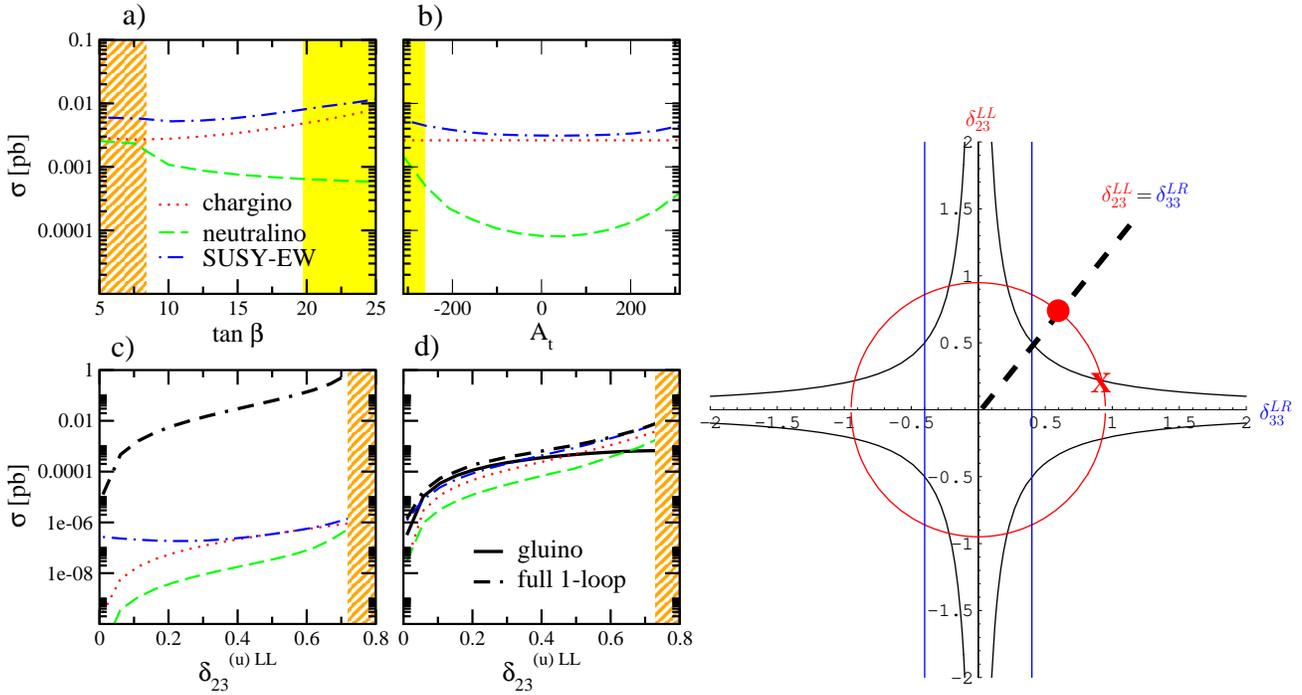}
\end{tabular}}
%
\caption{\footnotesize{(\textit{Upper-left panels}) SUSY-EW contribution to
$\sigma_{\Ptop\APcharm}$ (in $\picobarn$) as a function of
\textbf{a)} $\tan \beta$ and \textbf{b)} $A_t$. (\textit{Lower-left
panels}) Combined SUSY-QCD and SUSY-EW contributions to
$\sigma_{\Ptop\APcharm}$ (in $\picobarn$) as a function of
$\delta_{23}^{(u)\,LL}$ within \textbf{c)} a SUSY-QCD favored and
\textbf{d)} a SUSY-EW favored scenario. The shaded regions are
excluded by $\bsg$ restrictions, while the dashed ones are ruled out
by the mass bounds on the lightest supersymmetric particles.
(\textit{Right panel}) Sketch of the geometrical configuration
leading to the optimal SUSY-QCD contribution (see the text for
details).}} \vspace{-0.25cm} \label{fig:ew1}
\end{figure}

\vspace{0.1cm} Let us begin by adressing the SUSY-QCD contribution
to $\sigma(\processgg)$. In a general scenario within non-vanishing
flavor-mixing, the FCNC interactions are induced by
gluino-quark-squark vertices and lead to partonic cross sections of
the following (approximate) form \footnote{We use a standard
notation for the SUSY parameters, see e.g.\cite{our2} and references therein for
the concrete definitions.}:
\begin{equation}
\sigma (\gluglu) \sim\
\frac{|\delta_{23}^{(u)\,LL}|^2}{s}\,\left(\frac{\alpha_s^2}{16\,\pi^2}\right)^2\,
\frac{\mt^2 ({A_t}-\mu/\tb)^2}{\msusy^4} \label{eq:sqcd} \,.
\end{equation}
Let us observe that the same sort of gluino-mediated FCNC
interactions will also induce SUSY-QCD corrections to the low-energy
observable $\mathcal{B}_{exp}(\bsg)$ \cite{misiak}. Enforcing the predicted
$\mathcal{B}_{MSSM}(\bsg)$ to stay within the experimental bounds ($[2 -
4.5]\times 10^{-4}$ at the $3\sigma$ level \cite{babar}), we will
avoid artificially enhanced flavor-mixing sources. Moreover, due to
the $SU(2)_L$ gauge invariance, the  LL-blocks of the soft-SUSY
breaking squark masses get tied through $\left( M^{(u)\,LL}
\right)^2 = K\,\left(M^{(d)\,LL}\right)^2 K^\dagger $,
in such a way that the $\bsg$ bound on $\delta_{23}^{(d)\,LL}$
translates into a bound on $\delta_{23}^{(u)\,LL}$.
Our endeavor is to systematically seek for those (experimentally
allowed) regions in the MSSM parameter space where the SUSY-QCD
effects are maximum. We can manage to project this optimal regime by
means of an approximate analytical procedure. To begin with, let us
define $\delta_{33}^{(u)\,LR}=m_t\,(A_t-\mu/\tan\beta)/M^2_{SUSY}$, from
which we can rewrite Eq.~(\ref{eq:sqcd}) as $\sigma(\gluglu)=
(\delta_{23}^{(u)\,LL})^2\, (\delta_{33}^{(u)\,LR})^2\,/s$.
Notice that isolines of constant $\sigma=\sigma_0$ are hyperbolae in
the $\delta_{23}^{(u)\,LL}\times\delta_{33}^{(u)\,LR}$ plane. On the other
hand, upon diagonalization of the squark mass matrix, the
(approximate) value of the lightest squark mass reads
$m_{\tilde{q}_1}= M_{SUSY}^2 (1-\sqrt{(\delta_{23}^{(u)\,LL})^2 +
(\delta_{33}^{(u)\,LR})^2})$. The LEP mass bounds ($m_{\tilde{q}_1} \ge
90\,\GeV$) yield the inequality $(\delta_{23}^{(u)\,LL})^2 +
(\delta_{33}^{(u)\,LR})^2 < 1- m_{\tilde{q}_1}^2/ M_{SUSY}^2 \equiv
R^2$. We further impose that $|A_t|< 3M_{SUSY}$ in order to avoid
color-breaking minima, thus supplying a new restriction on
$\delta_{33}^{(u)\,LR}$. Now it is simply a matter of picking
out the point in the bisector line
$\delta_{33}^{(u)\,LR}=\delta_{23}^{(u)\,LL}$ where the outermost
hyperbola is tangent to the circle of radius $R$ (see
right-panel of Figure~\ref{fig:ew1}). The desired maximum, then, is:
\begin{equation}
 \delta_{23}^{(u)\,LL}=\frac{\sqrt{2}}{1+\left[1+\frac29\,
  m_{\tilde{q}_1}^2/{m_t}^2\right]^{1/2}}\simeq 0.7\,
\label{eq:max}\, ,
\end{equation}
\noindent from which we also find $A_t \simeq 2238\,\GeV$
and $M_{SUSY}\simeq 746\,\GeV$. A complete analysis of the behavior
of the cross section in the SUSY-QCD favored regime can be found in
\cite{our1,our2}. Therein it is shown that, for
relatively light gluino masses (of, say, $m_{\tilde{g}}\sim 200\,\GeV$)
the total cross section can reach $\sigma\sim 1\,\picobarn$, that
is, around $10^5$ $\Ptop\APcharm (\Pcharm\APtop)$ events per
$100\,\invfb$ of integrated luminosity. In view of the fact that the
SM contribution to this process reads $\sigma(\processgg) =
8.46\times10^{-8}\,\picobarn$, the maximum SUSY enhancement
is some $7$ orders of magnitude larger and could thus provide a
measurable signature at the $\picobarn$ level.

\vspace{0.1cm} The SUSY-EW contributions to $\processgg$ are
triggered by a large set of loop diagrams involving charginos,
neutralinos and charged Higgs bosons (see Ref.~\cite{our2} for
details). If gluinos are very heavy ($m_{\tilde{g}}\sim 2\,\TeV$),
or the FCNC parameter
$\delta_{23}^{(u)\,LL}$ very small, it may be interesting to explore the
behavior of $\sigma_{\tcbar}$ in a SUSY-EW favored scenario. Such a
regime can be achieved essentially by arranging a choice of MSSM
parameters that insures relatively light neutralinos, charginos
(within their respective LEP mass bounds, $m_{\tilde{\chi}^0_{1}}\ge
46\,\GeV, m_{\tilde{\chi}^{\pm}_{1}} \ge 96\,\GeV$), together with
light charged Higgs bosons and squarks (mainly stops), and of course in
compliance with the $\bsg$ restrictions. The resulting behavior of
$\sigma_{\tcbar}$ as a function of \textbf{a)} $\tan\beta$ and
\textbf{b)} $A_t$ is presented in the upper-left panels of
Figure~\ref{fig:ew1}.

In the lower-left panels of Figure~\ref{fig:ew1} we also quote the
$\sigma_{\tcbar}$ curves as a function of $\delta_{23}^{(u)\,LL}$ for
both \textbf{c)} SUSY-QCD favored and \textbf{d)} SUSY-EW favored
regimes. While in the former the gluino-mediated contribution
(with $m_{\tilde{g}} = 200\,\GeV$) is
largely dominant (namely, it lies at least $4$ orders of magnitude
over the chargino curve), with the highest production rates reaching
the level of $\sim 10^{5}$ events per $100\,\invfb$, the situation
is substantially different in the latter scenario. Therein we find
light neutralinos and charginos (close to their mass bounds) meeting
together with rather heavy ($m_{\tilde{g}}= 2\,\TeV$) gluinos. It is
remarkable that even for moderate values of $\delta_{23}^{(u)\,LL}$, the
SUSY-EW effects turn out to be significant, e.g. for
$\delta_{23}^{(u)\,LL} \sim 0.4$ the full one-loop curve is enhanced by a
factor $2$ with respect to the SUSY-QCD contribution. The larger
the flavor-mixing is, the larger the SUSY-EW effects become, achieving
maximal values that amount up to roughly
$\sigma=2\sigma_{\tcbar}\sim 0.01\,\picobarn$ ($1$ order of
magnitude above the SUSY-QCD contribution in this alternate
region of the parameter space).

\section{Discussion and conclusions}
We have discussed the supersymmetric production of single
top-quark final states at the LHC. Besides the fact that we only
allow explicit sources of flavor-mixing in the LL-block of the
squark mass matrix, the inclusion of the interplay between the
top-quark physics and the stringent low-energy $\bsg$ constraints,
provides a self-consistent framework to carry out the computation of
$\sigma(\processgg)=2\,\sigma_{\tcbar}$.
Related supersymmetric studies of SUSY-FCNC processes in the literature can
be found e.g. in \cite{tdecay,our1,our2,previs,santi} and references
therein. In this note we have revisited the SUSY-QCD
analysis of $\sigma_{\tcbar}$ presented in \cite{our1}, which was
extended by the inclusion of the SUSY-EW effects in \cite{our2}.
See also the latter reference for the potential impact of
the $\PB^0_s-\PaB^0_s$ mixing restrictions. Here we have singled out
only the regions of the MSSM parameter space where the SUSY-QCD
contribution becomes optimal, in full agreement with the stringent
and robust experimental restrictions imposed by $\bsg$. The most
favorable regimes are attained in the case of large
intergenerational mixing and relatively light gluinos, in which we
get cross sections up to $\sim 1\,\picobarn$ -- barely
$10^5$ events per $100\,\invfb$ of integrated luminosity.
Furthermore, we have also proved the existence of configurations
in the MSSM parameter space for which the SUSY-EW effects become
fully competitive with respect to the SUSY-QCD ones. It means that
for virtually decoupled gluinos (and so a largely supressed SUSY-QCD
contribution) the overall cross section may still be sizeable thanks
to the cooperative SUSY-EW flavor-changing interactions mediated by
charginos and neutralinos. Such regimes entail maximum cross
sections of the order of $\sim 0.01\,\picobarn$ - $1000$ events per
$100\,\invfb$. Worth noticing is that other alternative
sources of $\tcbar(\ctbar)$ final states in either the MSSM
\cite{santi} or the
Two-Higgs-doublet model \cite{thdm} are not competitive with the
direct SUSY mechanism $\processgg$. The latter might thus be, by
far, the most efficient mechanism of $\tcbar(\ctbar)$ FCNC
production within renormalizable extensions of the SM. 
Being the corresponding SM contribution
virtually absent, the presence of such $\tcbar(\ctbar)$ final
states, if effectively tagged and confidently isolated from the
background, could lead to evidence of supersymmetric physics.

\vspace{0.1cm}
\paragraph{Acknowledgments} DLV has been supported by the MEC FPU
grant Ref. AP2006-00357; JG and JS in part by MEC and FEDER under project
FPA2007-66665, and also by DURSI Generalitat de Catalunya under
project 2005SGR00564. JG is thankful to the Universidad de Zaragoza, for their kind
hospitality.

\newcommand{\JHEP}[3]{{\emph{JHEP}} {\bf #1} (#2)  {#3}}
\newcommand{\NPB}[3]{{ \emph{Nucl. Phys.} } { B\,{\bf#1}} ({ #2)}  {#3}}
\newcommand{\NPPS}[3]{{\sl Nucl. Phys. Proc. Supp. } {\bf#1} (#2)  {#3}}
\newcommand{\PRD}[3]{{ \emph{Phys. Rev.} } { D\,{\bf#1}} ( #2)   {#3}}
\newcommand{\PLB}[3]{{ \emph{Phys. Lett.} } { B\,{\bf#1}} (#2)  {#3}}
\newcommand{\EPJ}[3]{{ Eur. Phys. J } { C#1} (#2)  {#3}}
\newcommand{\PR}[3]{{ Phys. Rep } { #1} (#2)  {#3}}
\newcommand{\RMP}[3]{{ Rev. Mod. Phys. } { #1} (#2)  {#3}}
\newcommand{\IJMP}[3]{{ Int. J. of Mod. Phys. } { A#1} (#2)  {#3}}
\newcommand{\PRL}[3]{{ \emph{Phys. Rev. Lett.} } { {\bf #1}} (#2) {#3}}
\newcommand{\ZFP}[3]{{ Zeitsch. f. Physik } { C#1} (#2)  {#3}}
\newcommand{\IJMPA}[3]{{ Int. J. Mod. Phys. } { A#1} (#2) {#3}}
\newcommand{\MPLA}[3]{{ Mod. Phys. Lett. } {A#1} (19#2) {#3}}

\end{document}